\documentclass[12pt]{article}
\usepackage{amssymb}
\usepackage{amsmath}           

\begin{document}

\title{\bf NSVZ relation and the dimensional reduction in ${\cal N}=1$ SQED.}

\author{
S.~S.~Aleshin,\\
{\small{\em Institute for Information Transmission Problems}},\\ {\small{\em
Department of Quantum Physics,}}\\
{\small{\em 127051, Moscow, Russia,}}
}
\maketitle

\begin{abstract}
It is known that factorization of the $\beta$-function loop integrals into integrals of double total derivatives is an important ingredient needed for deriving the NSVZ relation by direct perturbative calculations in ${\cal N}=1$ SQED regularized by the higher derivatives. It allows to relate the $\beta$-function and the anomalous dimension of the matter superfields defined in terms of the bare coupling constant. In this work we find the analog of this result in the case of using dimensional reduction regularization in the lowest orders. However, we demonstrate that in this case the NSVZ relation is not satisfied for the RG functions defined in terms of the bare coupling constant. Nevertheless, it is possible to impose boundary conditions to the renormalization constants determining the NSVZ scheme in the three-loop order for the RG functions defined in terms of the renormalized coupling constant.
\end{abstract}

%
%

\unitlength=1cm

\section{Introduction}
$\phantom{aa}$\par
One of the most famous result in ${\cal N}=1$ abelian gauge theory (SQED) with $N_f$ flavors,
\begin{eqnarray}\label{1}
S=\frac{1}{4e_{0}^{2}}Re\int d^{4}xd^{2}\theta\, W^{a}W_{a}+
\sum_{i=1}^{N_{f}}\frac{1}{4}\int d^{4}xd^{4}\theta\,(\phi_{i}^{*}e^{2V}\phi_{i}+\tilde{\phi}_{i}^{*}e^{-2V}\tilde{\phi}_{i}),
\end{eqnarray}

\noindent is the relation between renormalization of the coupling constant and of the matter superfields (which is called the NSVZ $\beta$-function) \cite{Novikov:1983uc}:
\begin{eqnarray}\label{NSVZ}
\beta(\alpha_0)=\frac{\alpha_0^2N_{f}}{\pi}(1-\gamma(\alpha_0)).
\end{eqnarray}
Here $\alpha_0$ is the bare coupling constant and $\gamma(\alpha_0)$ is the anomalous dimension of the matter superfields. Eq. (\ref{NSVZ}) follows from the connection between two-point Green functions of the gauge and matter superfields which takes place in the case of using the higher derivative regularization \cite{Stepanyantz:2011jy,Stepanyantz:2014ima}. A part of the effective action connected with the two-point Green functions can be presented in the form:
\begin{eqnarray}
\Gamma^{(2)}-S_{gf}=-\frac{1}{16\pi}\int\frac{d^{4}p}{(2\pi)^{4}}d^{4}\theta\, V(-p,\theta)\partial^{2}\Pi_{1/2}V(p,\theta)d^{-1}(\alpha,\mu/p)\nonumber\\ +\frac{1}{4}\sum_{i=1}^{N_{f}}
\int\frac{d^{4}p}{(2\pi)^{4}}d^{4}\theta(\phi_{i}^{*}(-p,\theta)\phi_{i}(p,\theta)+\tilde{\phi}_{i}^{*}(-p,\theta)\tilde{\phi}_{i}(p,\theta))G(\alpha,\mu/p),
\end{eqnarray}
where $\alpha=\alpha(\mu)$ is the renormalized coupling constant and $\mu$ is the renormalization scale.

It was noted that loop integrals determining $\beta$-function in abelian supersymmetric theories can be presented as integrals of total derivatives \cite{Soloshenko:2003nc} and double total derivatives \cite{Smilga:2004zr} in the momentum space (in the limit of the vanishing external momentum) in the case of higher derivative regularization. This feature allows to reduce one of the loop integrals to an integral of the $\delta$-function:
\begin{eqnarray}\label{NSVZ_delta-function}
\frac{d}{d\ln\Lambda} \Big(d^{-1}
- \alpha_0^{-1}\Big)\Big|_{p=0}=\frac{d}{d\ln\Lambda} \Big({\text{One-loop}}-16\pi^{3}N_{f}\int\frac{d^{4}q}{(2\pi)^{4}}\delta^{4}(q)\ln G\Big).
\end{eqnarray}

\noindent This implies that one of the loop integrals can be explicitly calculated, and a $\beta$-function in $L$-loop approximation can be connected with an anomalous dimension of the matter superfields in the $(L-1)$-loop. As a consequence, the NSVZ $\beta$-function for ${\cal N}=1$ SQED with $N_f$ flavors for the RG functions defined in terms of the bare coupling constant can be obtained by summing supergraphs in the case of using the higher covariant derivative regularization.

Within the dimensional technique, the loop integrals cannot be presented in the form of integrals of double total derivatives. However, we will find the analogues of these structures in the theory regularized by the dimensional reduction at the three-loop level. We will find the influence of such structures on NSVZ relation and answer the question why the RG functions defined in terms of the renormalized coupling constant in the $\overline{\mbox{DR}}$-scheme do not satisfy the NSVZ relation. Besides, in the three-loop approximation we will find the boundary conditions to the renormalization constants giving the NSVZ scheme with the dimensional reduction for the RG functions defined in terms of the renormalized coupling constant.

\section{Structure of the three-loop scheme-dependent\\contribution to the $\beta$-function}

\qquad The expression for three-loop scheme-dependent contribution to the $d^{-1}$ function in ${\cal N}=1$ SQED with $N_f$ flavors regularized by the dimensional reduction can be presented in the form:
\begin{eqnarray}
&&\hspace*{-10mm} d^{-1}-\alpha_{0}^{-1}=8\pi N_{f}\Lambda^{\varepsilon}\int\frac{d^{d}k}{(2\pi)^{d}}\frac{1}{k^{2}(k+p)^{2}} -8\pi N_{f}\Lambda^{\varepsilon}\frac{\varepsilon}{1-\varepsilon}\int\frac{d^{d}k}{(2\pi)^{d}}\frac{1}{k^{2}(k+p)^{2}}(\ln G)_{\text{1--loop}}\nonumber\\
&&\hspace*{-10mm} -8\pi N_{f}\Lambda^{\varepsilon}\frac{2\varepsilon}{1-3\varepsilon/2}\int\frac{d^{d}k}{(2\pi)^{d}}\frac{1}{k^{2}(k+p)^{2}}(\ln G)_{\text{2--loops,$N_{f}$}}+\text{finite terms}+O(\alpha_{0}^{2}N_{f})+O(\alpha_{0}^{3})
\end{eqnarray}
where $\varepsilon\equiv 4-d$ and $\ln G_{\text{1-loop}}$ and $(\ln G)_{\text{2-loop,$N_f$}}$ are parts of the function $\ln G$ corresponding to the one- and a part of the two-loop approximation proportional to $N_f$, respectively.
From this expression we can conclude that two- and three-loop contributions to the $\beta$-function are related to the one- and two-loop contributions to the $\ln G$-function, respectively.
In the considered integrals the expressions \cite{Aleshin:2015qqc}
\begin{eqnarray}
\frac{1}{2\pi^{2}}\Lambda^{\varepsilon}\cdot\frac{(L-1)\varepsilon}{1-L\varepsilon/2}\int\frac{d^{d}q}{(2\pi)^{d}}\frac{1}{q^{2}(q+p)^{2}},
\end{eqnarray}
where $L$ is a number of loops, play the same role as the
\begin{eqnarray}
\int\frac{d^4q}{(2\pi^4)}\delta^4(q),
\end{eqnarray}
which appears after calculating integrals with higher derivatives regularization in the limit of the vanishing external momentum (see Eq.\eqref{NSVZ_delta-function}).

As a result, in the considered order we have received an analogue of this structure for dimensional reduction which explicitly connect two-point Green functions for the gauge and matter superfields as it makes integral of $\delta$-singularities in the theory regularized by higher derivatives.

\section{The NSVZ scheme in the three-loop order}

\qquad Let us try to derive the NSVZ-like relation for the RG functions defined in terms of the bare coupling constant:
\begin{eqnarray}\label{bare coupling constant}
\beta(\alpha_0)\equiv \frac{d\alpha_0}{d\ln\Lambda}\Big|_{\alpha=\mbox{\scriptsize const}};\qquad \gamma(\alpha_0) \equiv  - \frac{d\ln Z}{d\ln\Lambda}\Big|_{\alpha=\mbox{\scriptsize const}}.
\end{eqnarray}
Using the connection between two-point Green functions of gauge and matter
superfield described above we obtain that the $\beta$-function can be presented in NSVZ like form,
\begin{equation}
\frac{\beta(\alpha_0)}{\alpha_0^2} = \frac{N_f}{\pi}\Big(1 - \gamma(\alpha_0) \Big) + \frac{\Delta\beta(\alpha_0)}{\alpha_0^2}
+ O(\alpha_0^2 N_f) + O(\alpha_0^3),
\end{equation}
where
\begin{eqnarray}
&&\hspace*{-0.5cm} \Delta\beta(\alpha_0) = \frac{\alpha_0^4 (N_f)^2}{\pi^3}\cdot \lim\limits_{\varepsilon\to 0}
\Bigg(\frac{2}{\varepsilon}\Big[\frac{\varepsilon}{1-\varepsilon} \Big(\frac{4\pi\Lambda^2}{p^2}\Big)^{\varepsilon/2} G(1,1+\varepsilon/2) - 1\Big]  - \frac{\varepsilon}{4(1-\varepsilon)}\qquad\nonumber\\
&&\hspace*{-0.5cm}\times \Big(\frac{4\pi\Lambda^2}{p^2}\Big)^{\varepsilon} G(1,1) G(1,1+\varepsilon/2)  - \frac{1}{\varepsilon} \Big[\frac{3\varepsilon/2}{1-3\varepsilon/2} \Big(\frac{4\pi \Lambda^2}{p^2}\Big)^{\varepsilon/2} G(1,1+\varepsilon) -1\Big]\Bigg),\\
&&\qquad\qquad G(\alpha,\beta) \equiv \frac{\Gamma(\alpha+\beta-2+\varepsilon/2)}{\Gamma(\alpha)\Gamma(\beta)} B(2-\alpha-\varepsilon/2,2-\beta-\varepsilon/2).\nonumber
\end{eqnarray}
After some calculations, we obtain
\begin{equation}\label{Bare_Three-Loop_Beta}
\frac{\beta(\alpha_0)}{\alpha_0^2} = \frac{N_f}{\pi}\Big(1-\gamma(\alpha_0)\Big) + \frac{\alpha_0^2 (N_f)^2}{\pi^3}\Big(-\frac{1}{2\varepsilon} -\frac{1}{4}\Big) + O(\alpha_0^2 N_f) + O(\alpha_0^3).
\end{equation}
We see that, unlike the higher derivative regularization case, the RG functions defined in Eq.\eqref{bare coupling constant} do not satisfy the NSVZ relation, and the $\beta$-function explicitly depends on $\varepsilon$.

However, similarly to the higher derivative case \cite{Kataev:2013eta,Kataev:2013csa}, it is possible to impose boundary conditions determining the NSVZ-scheme with the dimensional reduction in the considered approximation \cite{Aleshin:2016rrr}:
\begin{equation}
\lim\limits_{\varepsilon\to\infty} \alpha_0(\alpha',\varepsilon, x_0=0)=\alpha' - \frac{\alpha'{}^3 N_f}{4\pi^2} + O(\alpha'{}^4);\qquad
\lim\limits_{\varepsilon\to\infty} Z'(\alpha',\varepsilon, x_0=0) =1.
\end{equation}

\section*{Acknowledgements}

The author is grateful K.V. Stepanyantz and A.L. Kataev for the useful discussions and valuable remarks. This work was performed at Institute for Information Transmission Problems with the financial support of the Russian Science Foundation (Grant No.14-50-00150)

\end{document}